\documentclass[12pt,letterpaper]{article}

\usepackage{graphicx}
\usepackage[body={17.5cm, 22cm},right=2cm]{geometry}
\usepackage{amssymb}
\usepackage{amsmath} 

\newcommand\ee{\end{equation}}
\newcommand\be{\begin{equation}}
\newcommand\eea{\end{eqnarray}}
\newcommand\bea{\begin{eqnarray}}
\newcommand{\sfrac}[2]{{\textstyle\frac{#1}{#2}}}
\newcommand\di{\partial}

\numberwithin{equation}{section}

\begin{document}

\begin{center}
\ 

\vspace{1.cm}

{\LARGE \bf{
Energy's and amplitudes' positivity}}\\[1cm]
{\large Alberto Nicolis$^{\rm a}$, Riccardo Rattazzi$^{\rm b}$, Enrico Trincherini$^{\rm c, d}$}
\\[.4cm]

\vspace{.2cm}
{\small \textit{$^{\rm a}$ 
Department of Physics and ISCAP,\\
 Columbia University, 
New York, NY 10027, USA }}

\vspace{.2cm}
{\small \textit{$^{\rm b}$ 
Institut de Th\'eorie des Ph\'enom\`enes Physiques, EPFL,\\
CHÐ1015 Lausanne, Switzerland}}        

\vspace{.2cm}
{\small \textit{$^{\rm c}$ 
Scuola Normale Superiore, \\
Piazza dei Cavalieri 7, 56126 Pisa, Italy}}

\vspace{.2cm}
{\small \textit{$^{\rm d}$
Scuola Internazionale di Studi Superiori Avanzati,  \\
Via Beirut 2-4, I-34151 Trieste, Italy}}

\end{center}

\vspace{.2cm}

\begin{abstract}

In QFT, the null energy condition (NEC) for a classical field configuration is usually associated with that configuration's stability against small perturbations, and with the sub-luminality of these. Here, we exhibit an effective
field theory that allows for stable NEC-violating solutions with exactly luminal excitations only.  The model is the recently introduced `galileon', or more precisely its conformally invariant version. 
We show that the theory's low-energy $S$-matrix obeys standard positivity as implied by dispersion relations.
However we also show that if the relevant NEC-violating solution is inside the effective theory, then {\em other} (generic) solutions allow for superluminal signal propagation.
While the usual association between sub-luminality and positivity is not obeyed by our example, that between NEC and sub-luminality is, albeit in a less direct way than usual.

\end{abstract}

%%%%%%%%%%%%%%%%%%%%%%%%%%%%%%%%%%%%%%%%%%%%%%%%%%%%%%%%%
%%%%%%%%%%%%%%%%%%%%%%%%%%%%%%%%%%%%%%%%%%%%%%%%%%%%%%%%%
\section{Introduction, and relation to previous work}

Of all energy conditions of general relativity (GR), the `null' one stands out as the best defined---thus the most likely to follow from fundamental properties of well-behaved physical systems. The reason is that the null energy condition (NEC),
\be
T_{\mu\nu} \, n^\mu n^\nu \ge 0 \qquad \mbox{for all null $n^\mu $'s} \; ,
\ee
is saturated by the cosmological constant contribution's to $T_{\mu\nu}$. It is therefore impossible to violate the NEC by adding a suitably negative c.c.~to an healthy system, or conversely, to obey it by adding a large positive c.c.~to an otherwise negative-energy system. This is not the case for other energy conditions. One could postulate that the latter be obeyed by the ``non-cosmological constant part'' of $T_{\mu\nu}$, but the splitting between the cosmological constant and everything else is at best non-local in field space and thus in real space---in the sense that it cannot be performed on the basis of local observations---and  more likely ill-defined.

%%%%%%%%%%%%%%%%%%%%%%%%%%%%%%%%%%%%%%%%%%%%%%%%%%%%%%%%%
\subsection{A no-go theorem\dots}
Since the value of the c.c.~does not affect the behavior of a theory at distances smaller than the curvature radius
%---and at all if we decouple gravity---
it is on the NEC that we should focus to look for a generic relation with good properties of healthy physical systems; moreover because asking if the stress-energy tensor of a given system violates the NEC is a well-defined question at scales where gravity is negligible, in the rest of the paper we will consider the limit of decoupled gravity.   

It is notoriously difficult to violate the NEC while retaining stability. Indeed, it has been shown in ref.~\cite{DGNR} that for a generic effective theory of scalar fields $\phi^I$ 
\footnote{Such a system is quite general. For instance it describes the low-energy dynamics of perfect fluids and solids \cite{DGNR}.
}, the NEC follows from stability and isotropy, or alternatively from stability and sub-luminality. Explicitly, given the low-energy effective Lagrangian as a derivative expansion,
\be
{\cal L} = {\cal L} \big( \phi^I, \di \phi^I , \di \di \phi^I , \dots \big) \; ,
\ee
at low energies one can restrict to the lowest-derivative order---at most one derivative per field---and ignore higher derivative terms,
\be
{\cal L} \to {\cal L } \big( \phi^I, B^{IJ} \big) \; , \qquad B^{IJ} \equiv \di_\mu \phi^I \, \di^\mu \phi^J \; . 
\ee
Then one can relate the derivatives of the Lagrangian w.r.t.~the metric to the derivatives of the Lagrangian w.r.t.~the scalars' derivatives. That is, there is a very direct relation between a generic solution's stress-energy tensor, and the kinetic action for small perturbations of the same solution.
One can thus show that if a solution violates the NEC, either {\em (i)} it is unstable, or {\em (ii)} if it is stable, both its stress-energy tensor is anistropic, and some of its perturbations are super-luminal.
Conversely, if a solution is stable and has an isotropic stress-energy tensor, or if it stable and admits sub-luminal
excitations only, then it obeys the NEC \cite{DGNR}.

The instabilities we are concerned about are those associated with negative kinetic or gradient energy for the fluctuations, rather than with negative mass terms
\footnote{For background solutions that break Lorentz-invariance spontaneously, the definition of stability may be  subtler. See ref.~\cite{DGNR} for a discussion about this point.}.
This is because such instabilities plague all short-wavelength fluctuations, down to the UV cutoff of the theory. The decay rate is thus extremely fast, dominated by the shortest scales in the theory, and cannot be reliably computed within the effective theory. Conversely, a standard tachyon-like instability, like Jeans's or that associated with the negative mode of a true-vacuum bubble, is dominated by modes with momenta of order of the tachyonic mass scale, which can be parameterically smaller than the UV cutoff.

Now, the theorem summarized above crucially relies on neglecting higher-derivative contributions to the Lagrangian. Is this justified?
Usually it is, for higher-derivative terms are irrelevant at low energies and, perhaps more to the point, if they do become important the derivative expansion is expected to break down---with an infinite tower of even higher-order terms all becoming important at the same time. At the level of the classical e.o.m., higher-derivative terms signal the presence of massive ghost-like degrees of freedom, thus {\em demanding} that the effective theory break down at energies below their masses and effectively making the higher-derivative terms unimportant within the regime of validity of the effective theory.

%%%%%%%%%%%%%%%%%%%%%%%%%%%%%%%%%%%%%%%%%%%%%%%%%%%%%%%%%
\subsection{\dots with exceptions}
However there can be exceptions to this general logic.
The first exception is provided by the ghost-condensate \cite{ACLM}. There, the fluctuations' Lagrangian  is degenerate (or nearly so) at the lowest-derivative level, and as a consequence certain higher-derivative terms can be important at low energies, below the cutoff of the theory. This is consistent with the derivative expansion because the background solution---the `condensate'---spontaneously breaks Lorentz invariance. It is then higher {\em spatial} derivative terms that become important for the fluctuations' dynamics. They do not introduce any new ghost-like degrees of freedom at the classical level, and being the leading source of gradient energy, they change the scaling dimensions of higher-order operators thus keeping the derivative expansion under control \cite{ACLM}. Indeed the ghost-condensate was used in ref.~\cite{CLNS} to construct a cosmological solution that violates the NEC while being stable and isotropic, and  having (highly) sub-luminal excitations only.

Here we consider a second exception---that provided by the galileon \cite{NRT}. The model involves a scalar $\pi$ enjoying a shift-symmetry on itself as well as on its first derivatives,
\be \label{galilean}
\pi \to \pi + c + b_\mu x^\mu \; ,
\ee
with constant $c$ and $b_\mu$.
This is reminiscent of the invariance of non-relativistic mechanics under galilean transformations $x(t) \to x(t) + x_0 + v_0  t  $---hence the name of the model.
Now, galilean invariance (\ref{galilean}) forces each $\pi$ in the e.o.m.~to be acted upon by {\em at least} two derivatives. Absence of ghosts requires instead that there be {\em no more} than two derivatives per field. There is therefore a preferred set of higher-derivative terms in the Lagrangian---those that yield {\em purely} two-derivative equations of motion for $\pi$. Such terms have been classified in ref.~\cite{NRT}. In four spacetime dimensions there are four: the standard kinetic term
\be \label{L2}
{\cal L}_2 = -\sfrac12 (\di \pi)^2 \; ,
\ee
a cubic interaction
\be \label{L3}
{\cal L}_3 = -\sfrac12 (\di \pi)^2  \Box\pi \; ,
\ee
and quartic and quintic ones
\be \label{L45}
{\cal L}_4 \sim \di \pi \di \pi (\di^2 \pi)^2 \; , \qquad {\cal L}_5 \sim \di \pi \di \pi (\di^2 \pi)^3 \; ,
\ee
with suitable Lorentz contractions
\footnote{Formally, the tadpole $\pi$ should also be added  to this class \cite{NRT}. However in its presence the trivial configuration $\pi=0$ is not a solution. We will therefore set it to zero.}.

Now, if we stick to this particular set of interactions,
at the classical level no new ghost-like degree
of freedom shows up in the non-linear regime---by construction. This is a necessary condition for the quantum
effective theory not to break down for sizable classical non-linearities. In fact the model exhibits
remarkable properties at the quantum level as well. First, thanks to galilean invariance, the Lagrangian terms we discussed do not get renormalized upon inclusion of quantum corrections, nor do the latter generate terms with fewer derivatives, like e.g.~$(\di \pi)^4$ \cite{LPR}.
Second, with a consistent,
if not generic choice of counter-terms, our effective theory can be extrapolated deep into the
non-linear regime of classical solutions. Essentially, galilean invariance allows for a robust disentaglement of large classical non-linearities from large quantum effects. See ref.~\cite{NR} for details and discussions about this point.

In summary, the galileon provides us with a classically as well as quantum-mechanically consistent higher-derivative Lagrangian, which can potentially evade the no-go theorem reviewed above. Analyzing this possibility is the goal of our paper.

%%%%%%%%%%%%%%%%%%%%%%%%%%%%%%%%%%%%%%%%%%%%%%%%%%%%%%%%%
%%%%%%%%%%%%%%%%%%%%%%%%%%%%%%%%%%%%%%%%%%%%%%%%%%%%%%%%%
\section{The NEC-violating solution}\label{NECsolution}

For practical purposes and for reasons that will become clear below, it is more convenient to work with a slight deformation of the theory. We can promote the Poincar\'e group supplemented by the galilean transformations (\ref{galilean}) (a fifteen-dimensional symmetry group overall) to the conformal group $SO(4,2)$ \cite{NRT}. Indeed, if we think of the conformal group as the subgroup of diffeomorphisms that leaves the fictitious metric
\be \label{metric_pi}
g^{(\pi)}_{\mu\nu} (x) \equiv e^{2 \pi(x)} \eta_{\mu\nu} 
\ee
conformally flat, then we recognize in eq.~(\ref{galilean}) the small-$\pi$ limit of infinitesimal dilations
\be \label{dilation}
\pi(x) \to \pi (  \lambda  x ) + \log \lambda \; ,
\ee
and of infinitesimal special conformal transformations
\be \label{special}
\pi(x) \to \pi \big( x + (b \, x^2 - 2 (b \cdot x) x)  \big) - 2 b_\mu x^\mu \; .
\ee
Then one can show that for each galilean invariant term, there is a corresponding conformally invariant one that reduces to that for small $\pi$. 
For instance the kinetic and cubic terms (\ref{L2}, \ref{L3}) get promoted to
\bea
{\cal L}_2 & \to & -\sfrac12 \, e^{2\pi} (\di \pi)^2  \label{L2c}\\
{\cal L}_3 & \to &  -\sfrac12 (\di \pi)^2  \Box\pi - \sfrac14 (\di \pi)^4 \label{L3c}\; .
\eea
Such conformally invariant `completion' is most easily constructed by taking  suitable curvature invariants for the metric (\ref{metric_pi}) in terms of the Ricci tensor $R_{\mu\nu}$, as described in ref.~\cite{NRT}. Notice that the Riemann tensor does not appear because it can be rewritten in terms of the Ricci since the metric (\ref{metric_pi}) is conformally flat.     
Working with the model's conformal version makes finding interesting highly symmetric solutions and studying their properties particularly easy. For instance, there is a deSitter-like solution where the fictitious metric (\ref{metric_pi}) describes (part of) deSitter space:
\be \label{dS_solution}
e^{\pi_{\rm dS}} = - \frac{1}{H_0 t} \; , \qquad -\infty < t < 0 \; .
\ee
Such a solution spontaneously breaks the conformal group $SO(4,2)$ down to one of its maximal subgroups, the deSitter group $SO(4,1)$. 
The `curvature' scale $H_0$ depends on the model's parameters. Starting from a generic linear combination
of our Lagrangian terms
\be
{\cal L} = c_2 {\cal L}_2 + \dots +c_5  {\cal L}_5 \; ,
 \ee
where with an abuse of notation by ${\cal L}_n$ we refer to the conformal versions of the galilean invariants (\ref{L2}--\ref{L45}), $H_0$ is a solution to the algebraic equation \cite{NRT}
\be \label{dS_eom}
H_0^2 \big( -2 c_2  +3 c_3 \, H_0^2 -3 c_4 \, H_0^4 + \sfrac32 c_5 \,  H_0^6 \big) = 0 \; .
\ee
Notice that, since we work with the dimensionless field $\pi$, the lagrangian coefficients are dimensionful: $c_n\propto E^{6 -2n}$.  
To make conformal invariance explicit,  one could conveniently define a dimensionful dilaton field $\phi\equiv M e^\pi$, where $M$ is an arbitrary mass scale, and write the lagrangian in terms of $\phi$ and dimensionless couplings $\bar c_n=c_n M^{2n-6}$. The action parameterized this way would not depend explicitly on $M$, thanks to scale invariance.  In section 3 we shall conveniently pick $M\equiv H_0$. Here we prefer to use the
somewhat redundant notation with dimensionful $c_n$  for basically two reasons. First we want to make contact with the non-conformal galileon limit. Second, the role of conformal invariance in our dicussion is mostly technical: to simplify computations and to deliver a simple globally defined solution around which to expand. The lesson about NEC violation is more general.

Now, let us consider fluctuations about eq.~(\ref{dS_solution}). They are exactly luminal, because of the high degree of symmetry: by construction the full action only involves the fictitious metric (\ref{metric_pi}) and its curvature invariants. For the background solution we are expanding about, the latter are proportional to the former, since the background is maximally symmetric. Therefore, although the solution (\ref{dS_solution}) spontaneously breaks Lorentz invariance, the only background tensor structure appearing in the fluctations' action is $\eta_{\mu\nu}$. In particular, it is easy to see that the quadratic Lagrangian for the fluctuations $\varphi \equiv \pi-\pi_{\rm dS}$ is
\be \label{fluctuations}
{\cal L}_\varphi = d_2 \frac{1}{H_0^4 t^4} \left[ - \sfrac12 H_0^2 t^2 (\partial \varphi)^2 - \sfrac12 (- 4 H_0^2) \varphi^2 \right] \; ,
\ee
where the overall normalization factor $d_2$ is \cite{NRT}
\be  \label{d2}
d_2 \equiv c_2 - 3 c_3 \, H_0^2 +\sfrac92 c_4 \, H_0^4 - 3 c_5 \, H_0^6 \; .
\ee
The kinetic term is fixed by the above symmetry considerations, and its normalization by the local analysis of ref.~\cite{NRT}. The mass term is also enforced by symmetry. The background $\pi$ configuration spontaneously break time-translational invariance, $t \to t + \epsilon$. The fluctuation $\varphi$ is the associated Goldstone boson, which has to non-linearly represent the broken symmetry:
\be
\varphi \to \varphi  + \pi'_{\rm dS} (t) \cdot \epsilon = \varphi - \frac{\epsilon}{t} \; .
\ee
The variation of the kinetic term under an infinitesimal transformation is non-zero:
\be
\delta\big( -\sfrac12 \sfrac{1}{H_0^2 t^2}(\partial \varphi)^2 \big) = \frac{1}{H_0^2 t^2} \,  \partial_\mu \varphi \, \partial^\mu \frac{\epsilon}{t} = \epsilon \cdot 4 \varphi \,  \frac{1}{H_0^2 t^5} \; ,
\ee
where in the last step we integrated by parts. Then the kinetic term has to be accompanied by a mass term that compensates for this variation,
\be
\delta\big( -\sfrac12 \sfrac{1}{H_0^4 t^4} m ^2 \varphi^2 \big) =\epsilon \cdot m^2 \varphi \, \frac{1}{H_0^4 t^5}  \; ,
\ee
which happens precisely for $m^2 = - 4 H_0^2$.
This is obviously the same logic that for spontaneously broken {\em internal} symmetries ensures the masslessness of the associated Goldstone bosons.

Notice that while the fluctuation's kinetic energy is positive definite for
\be \label{dS_stability}
d_2 > 0 \; ,
\ee
the mass term always has the wrong sign. However its size is of order of the typical time variation rate of the background solution. As a consequence ascertaining the solution's stability is subtler than for a time-independent one.
In fact the very concept of stability is somewhat ambiguous for a time-dependent solution (see a related discussion in ref.~\cite{NR}). As to short wave-length perturbations, the solution is certainly stable, for the fluctuations' kinetic energy is positive, and the mass term is negligible at high frequencies. For longer wavelengths, we can tentatively define stability in the following way. For given initial conditions, the fluctuation's equation of motion yields two solutions.
At zero momentum, one of these will always be the Goldstone solution:
\be \label{goldstone}
\varphi_{\rm Goldstone} \equiv \pi'_{\rm dS} (t) \cdot \epsilon =  - \frac{\epsilon}{t} \; ,
\ee
with constant $\epsilon$. {\em If} this is the leading solution at late times ($t \to 0^-$, in our case), we say that the background solution $\pi_{\rm dS}$ is stable. Indeed in such a case any long-wavelength small departures from the background solution will be diluted away at sufficiently late times, and we will be left with the background solution itself, slightly translated in time. This is hardly an instability. It is straightforward to check that for the Lagrangian (\ref{fluctuations}) the Goldstone solution (\ref{goldstone}) is indeed the leading one at late times.

In summary, the deSitter solution is stable against small perturbations if $d_2$ defined in eq.~(\ref{d2}) is positive. Also perturbations are exactly luminal, thanks to the solution's high degree of symmetry.

%
%fluctuations live in a fictitious deSitter metric, with the real-space Minkowski coordinates $x^\mu = (t, \vec x)$ playing the role of conformal FRW coordinates,
%\be
%g^{(\pi)}_{\mu\nu} (x)  \, dx^\mu d x^\nu = \frac{1}{H_0^2 t^2} (-dt^2 + d \vec x \, ^2) \; ,
%\ee
%for  which the light-cone has aperture one. 

%%%%%%%%%%%%%%%%%%%%%%%%%%%%%%%%%%%%%%%%%%%%%%%%%%%%%%%%%
\subsection{The stress-tensor}
It remains to be determined whether such a solution can violate the NEC. To answer this question, we have to compute the stress-energy tensor,
\be
T^{\mu\nu} = \frac{2}{\sqrt{-g}} \frac{\delta S_\pi}{\delta g_{\mu\nu}} \; .
\ee
For a generic $\pi$ configuration the computation is straightforward, but painful due to the higher-derivative structure of the Lagrangian.
However for our deSitter solution (\ref{dS_solution}) things drastically simplify, as we now show. The reader not interested in the details of the computation can jump directly to eq. (\ref{NECv}), which expresses the condition on the coefficients $c_i$ for the solution to violate the NEC.

All conformal-invariant operators are of the form \cite{NRT}
\be
e^{(4-2n) \pi} (\di^2 \pi)^m (\di \pi)^{2n - 2m} \sim \frac1{t^4}
\ee
This is just a consequence of scale-invariance, which is unbroken by the solution:
\be
e^{2\pi_{\rm dS}(t)} \to e^{2\pi_{\rm dS}(\lambda t) + 2 \log \lambda} =  e^{2\pi_{\rm dS}(t)}
\ee
Therefore
\be
T^{\mu\nu} = \tau^{\mu\nu} \frac{1}{t^4} \;,
\ee
with a constant and isotropic $\tau^{\mu\nu}$.
Now, conservation of $T^{\mu\nu}$ for an homogeneous solution in flat space is simply
\be
\frac{d \rho}{d t} = 0 \; ,
\ee
which implies $\tau^{00}=0$. We are left with the pressure $T^{ij} = p \, \delta^{ij}$, which we can get from the trace
\be
3p = T^{\mu} {}_\mu \; ,
\ee
which in turn we can compute by varying the action just w.r.t.~the conformal mode of the metric:
\be
T^{\mu} {}_\mu = \eta_{\mu\nu} \, 2 \left. \frac{\delta S_\pi}{\delta g_{\mu\nu}} \right |_{\eta_{\mu\nu}} =\left.  \frac{\delta S_\pi}{\delta \varphi}\right |_{\varphi=0}  \; ,
\ee
where
\be \label{metric_phi}
g_{\mu\nu} (x)= e^{2\varphi(x)} \eta_{\mu\nu} \equiv g^{(\varphi)}_{\mu\nu} (x) \; .
\ee

Now we can use the fact that $S_{\pi}$ is built out of curvature invariants involving the fictitious metric (\ref{metric_pi}),
where $\pi$ itself appears as the conformal mode. More precisely, suppose we compute the Ricci tensor for the metric $e^{2 \pi} g_{\mu\nu}$, with $g_{\mu\nu}$ generic. We get
\be
R_{\mu\nu} [e^{2 \pi} g ] = R_{\mu\nu} [g] + R_{\mu\nu} [e^{2 \pi} \eta, g ] \; ,
\ee
where by the last term we denote the Ricci tensor for the metric $(\ref{metric_pi})$, ``covariantized'' with the metric $g_{\mu\nu}$---precisely what makes up $S_\pi$ upon minimal coupling. Actually, to perform the contractions appearing in $S_\pi$ we have to raise one index with the metric $e^{2 \pi} g_{\mu\nu}$:
\be
R^\mu {}_\nu [e^{2 \pi} g ] = e^{-2\pi} R^\mu {}_\nu [g] + e^{-2\pi} g^{\mu\alpha}  R_{\alpha\nu} [e^{2 \pi} \eta, g ] \; ,
\ee
where by $R^\mu {}_\nu [g]$ we are denoting $g^{\mu\alpha} R_{\alpha\nu} [g]$. As we mentioned, minimal coupling corresponds to the replacement
\bea
S_\pi  =  \int \! d^4x \, e^{4\pi}  F \big( R^\mu {}_{\nu} [e^{2\pi} \eta] \big ) \nonumber 
&\to & \int \! d^4x \, \sqrt{-g} \,  e^{4\pi}  F \big( e^{-2\pi} g^{\mu\alpha}  R_{\alpha\nu} [e^{2 \pi} \eta, g ] \big ) \nonumber  \\
& = & \int \! d^4x \, \sqrt{-g} \, e^{4\pi}  F \big( R^\mu {}_\nu [e^{2 \pi} g ] - e^{-2\pi} R^\mu {}_\nu [g]  \big )
\eea
which for the conformally flat metric (\ref{metric_phi}), at linear order in $\varphi$ becomes
\bea
S_\pi [\pi, \varphi] & = & \int \! d^4x \, e^{4(\pi+\varphi)}  F \big( R^\mu {}_\nu [e^{2 (\pi+\varphi)} \eta ] - e^{-2\pi} R^\mu {}_\nu [e^{2\varphi} \eta]  \big )  \label{S_pi_phi_1}  \nonumber \\
& \simeq & S_\pi [\pi+\varphi, 0] - \int \! d^4x \, e^{2 \pi} \frac{\partial F}{\partial R^\mu {}_\nu} R^\mu {}_\nu [e^{2\varphi} \eta] \nonumber \\
& \simeq & S_\pi [\pi+\varphi, 0]  + \int \! d^4x \, \varphi(x) \, \big( 2 \, \di^\mu \di_\nu + \delta^\mu_\nu \Box \big) 
\bigg( e^{2\pi} \frac{\partial F}{\partial R^\mu {}_\nu}\bigg)          \label{S_pi_phi_3}
\eea
On a solution of the $\pi$ eom, the first term has vanishing variation w.r.t.~$\varphi$, and we are left just with the second:
\be \label{trace}
T^{\mu} {}_\mu = \big( 2 \, \di^\mu \di_\nu + \delta^\mu_\nu \Box \big) 
\bigg( e^{2\pi} \frac{\partial F}{\partial R^\mu {}_\nu}\bigg) \; .
\ee

The above expression holds for any solution. However in our deSitter case we can simplify this expression even further.
For the deSitter solution,
\be
{R_{{\rm dS}}}^\mu {}_\nu = 3 H_0^2 \delta^\mu_\nu \; .
\ee
It is straightforward to convince oneself that
\be 
\left. \frac{\partial F}{\partial R^\mu {}_\nu} \right |_{\rm dS} = \frac{1}{12} \, \delta^\mu_\nu \,  \frac{d F({R_{{\rm dS}}}^\mu {}_\nu)}{d H_0^2}
\ee
Then derivatives in eq.~(\ref{trace}) only act on the $e^{2\pi}$ factor, and we finally get
\be \label{trace_final}
T^\mu {}_\mu = - \frac{3}{H_0^2 t^4} \,\frac{d F({R_{{\rm dS}}}^\mu {}_\nu)}{d H_0^2}
\ee
In summary, to compute the pressure, we just have to express the action of $\pi$ in terms of curvature invariants of the metric (\ref{metric_pi}), plug in the deSitter solution, and then derive w.r.t.~to $H_0^2$.

%%%%%%%%%%%%%%%%%%%%%%%%%%%%%%%%%%%%%%%%%%%%%%%%%%%%%%%%%
\subsection{The problem with ${\cal L}_3$}
If we apply eq.~(\ref{trace_final}) to  ${\cal L}_3$, we run into a problem. In order to express ${\cal L}_3$ in terms of curvature invariants, on top of $R^2$ and $ (R^\mu {}_{\nu})^2$ we need the operator \cite{NRT}
\be \label{generic_d}
\int \! d^d x \sqrt{-g_d} \, \frac{1}{d-4} \left[ \frac{(R^\mu {}_\nu)^2 }{d-1} - \frac{R^2}{(d-1)^2} \right]  \; ,
\ee
which diverges when $d \to 4$ for generic metric, but is regular for conformally flat metrics $g_{\mu\nu} = e^{2 \pi} \eta_{\mu\nu}$, for which it reduces to
\be \label{cubic_d}
\int \! d^d x \, e^{(d-4) \pi} \Big[ \sfrac{(d-2)(3d-4)}{2(d-1)} \,  \Box \pi (\di \pi)^2 + \sfrac{(d-2)^3}{2(d-1)} \,  (\di \pi)^4 + (\Box \pi)^2 \Big] \; .
\ee
Now, the problem is that to get from eq.~(\ref{generic_d}) to eq.~(\ref{cubic_d}), we integrated by parts. This amounts to neglecting a boundary term, whose coefficient diverges for $d \to 4$. This is fine for asymptotically trivial $\pi$'s, but for our deSitter solution we have to keep track of the boundary term. Indeed eq.~(\ref{generic_d}) diverges when computed on a deSitter solution, even though the latter {\em is} conformally flat. The bottom line is that if we want to apply eq.~(\ref{trace_final}) to an action containing ${\cal L}_3$, we must be more careful about boundary terms.

We can bypass this problem by treating ${\cal L}_3$ separately, since the computation of its stress-energy tensor is straightforward anyway:
\bea
\int \! d^4 x \, {\cal L}_3 \quad \to \quad  T^{(3)}_{\mu\nu} &  = & 
\sfrac12 \big[ 2 \, \di_\mu \pi \di_\nu \pi  \Box \pi - \big(\di_\mu \pi  \, \di_\nu ( \di \pi)^2 + \di_\nu \pi \,  \di_\mu ( \di \pi)^2 \big)
+ \eta_{\mu\nu} \, \di_\alpha \pi \,   \di^\alpha (\di \pi)^2
\big] \nonumber  \\
& + &\sfrac14 \big[ 4 ( \di \pi)^2 \di_\mu \pi \di_\nu \pi -  \eta_{\mu\nu} (\di \pi)^4 
		\big] \; .
\eea
[In computing the first line, which comes from eq.~(\ref{L3}), care has to be taken in varying the $\Box$, because it contains a covariant derivative. However this is straightforward to do if we rewrite $\Box \pi$ as $\frac{1}{\sqrt{-g}} \di_{\mu} (\sqrt{-  g} \,  g^{\mu\nu} \di_\nu \pi)$.]\\
In particular, the trace $T^{(3) \, \mu} {}_\mu$ computed on the deSitter solution $e ^{2 \pi_{\rm dS}} = 1/H_0^2 t^2$ is
\be
T^{(3) \, \mu} {}_\mu = \frac{3}{t^4} \; .
\ee
Now we can subtract the ${\cal L}_3$ contribution from the action, and apply eq.~(\ref{S_pi_phi_3}) to the remainder. The first term in eq.~(\ref{S_pi_phi_3}), which would vanish on-shell for the full action, now can be rewritten as minus the e.o.m.~contribution coming from ${\cal L}_3$. Combining everything we thus get
\be
T^{\mu} {}_\mu = c_3 \, T^{(3) \, \mu} {}_\mu  - c_3 \frac{\delta {\cal L}_3}{\delta \pi} 
- \frac{3}{H_0^2 t^4} \,\frac{d F^{(-3)} ({R_{{\rm dS}}}^\mu {}_\nu)}{d H_0^2}   \; ,
\ee
where the last term is the same as eq.~(\ref{trace_final}), applied to the non-${\cal L}_3$ part of the action.
As to second term, for the deSitter solution it is
\be
- c_3 \frac{\delta {\cal L}_3}{\delta \pi}  = - c_3 \big[(\Box \pi)^2 - \di_\mu\di_\nu \pi \, \di^\mu \di^\nu \pi + \di^\mu \big(\di_\mu \pi(\di \pi)^2 \big)  \big] = -c_3 \frac{3}{t^4} \; ,
\ee
which exactly cancels the first term.
In conclusion, eq.~(\ref{trace_final}) holds on-shell if we restrict it to the non-${\cal L}_3$ part of the action:
\be
T^\mu {}_\mu = - \frac{3}{H_0^2 t^4} \,\frac{d F^{(-3)}({R_{{\rm dS}}}^\mu {}_\nu)}{d H_0^2}
\ee

%%%%%%%%%%%%%%%%%%%%%%%%%%%%%%%%%%%%%%%%%%%%%%%%%%%%%%%%%
\subsection{The recipe}\label{recipe}
We are now in a position to check whether there exists a choice for the Lagrangian coefficients $c_2, \dots, c_5$ that makes our deSitter solution (\ref{dS_solution}) violate the NEC without giving up stability. (As we stressed, absence of superluminality  is automatic for this particular solution.)
The various contributions to
\be
F^{(-3)} = c_2 {\cal L}_2 + c_4 {\cal L}_4 + c_5 {\cal L}_5
\ee
computed on the deSitter solution are \cite{NRT}
\bea
{\cal L}_2 & = & -\sfrac1{12} R = -H_0^2 \\
{\cal L}_4 &= & - \sfrac1{8} \left[- \sfrac{7}{36} \, R^3 +  R (R^{\mu} {}_\nu)^2
- (R^\mu {}_\nu)^3 \right] =  \sfrac{3}{2} H_0^6 \label{culprit}\\
{\cal L}_5 &= &  - \sfrac1{4} \left[ \sfrac{93}{2 \cdot 6^4} \, R^4 - \sfrac{39}{4 \cdot 6^2} \, R^2 (R^\mu {}_\nu)^2 + \sfrac5{12}  \, R (R^\mu {}_\nu)^3 + \sfrac3{16} \big((R^\mu {}_\nu)^2\big)^2 -\sfrac{3}{8} (R^\mu {}_\nu)^4 
\right] =  -\sfrac{3}{8} H_0^8  
\eea
We thus get
\bea
p = \sfrac13 T^\mu {}_\mu & = &  -\frac{1}{H_0^2 t^4} \big[ -c_2 + \sfrac{9}{2} c_4  \, H_0^4- \sfrac{3}{2} c_5 \, H_0^6 \big]  \\
& =  &  -\frac{1}{H_0^2 t^4} \big[-3 c_2 +3 c_3 \, H_0^2 + \sfrac{3}{2} c_4 \, H_0^4 \big]  \; ,
\eea
where in the second step we used the e.o.m.~(\ref{dS_eom}).
To violate the NEC, since the energy density vanishes on the de Sitter solution, it is enough to have negative pressure, that is
\be \label{NECv}
-3 c_2  +3 c_3 \, H_0^2 + \sfrac{3}{2} c_4 \, H_0^4  > 0 \; .
\ee
On the other hand, for the de Sitter solution to be stable, we need (\ref{dS_stability}), that is
\be \label{positived2}
-3 c_2 + 3 c_3 \, H_0^2 -\sfrac32 c_4 \, H_0^4  >  0 \; ,
\ee
where we used the e.o.m.~(\ref{dS_eom}).
Notice that  the e.o.m.~(\ref{dS_eom}) is the only one of our conditions where $c_5$ appears. We can thus think of it as determining $c_5$ for given $H_0^2$ and $c_2, c_3, c_4$. Furthermore, we can eliminate $H_0^2$ by absorbing it into a redefinition of $c_2, c_3, c_4$. That is, for the moment we can work in $H_0^2 = 1$ units and reintroduce $H_0^2$ later by dimensional analysis.
Then, combining eqs.~(\ref{NECv}) and (\ref{positived2}) we get a condition on $c_4$:
\be\label{c4}
- 2 (c_3 - c_2)  < c_4 <   2 (c_3 - c_2) \; .
\ee
This interval exists if and only if
\be \label{c3}
c_3 > c_2  \; .
\ee
So, summarizing, here is the recipe for a system that has a stable NEC-violating, de Sitter-like solution:
\begin{enumerate}
\item Pick $H_0^2>0$ for the deSitter-like solution;
\item Pick $c_2$; 
\item Pick $c_3  >  c_2/H_0^2 \;$;
\item Pick $c_4$ in the interval $ - 2 (c_3 H_0^2 - c_2)  < c_4 H_0^4 <   2 (c_3 H_0^2 - c_2) $;
\item $c_5$ is fixed by the e.o.m.: $c_5 \, H_0^6 = -\sfrac23(-2c_2 +3 c_3\, H_0^2 - 3 c_4 \, H_0^4) \; $;
\item The desired action is $S_\pi = \int d^4x \, c_2 {\cal L}_2 + \dots + c_5 {\cal L}_5 \; $.
\end{enumerate}
Notice that if we want the trivial configuration $\pi=0$ to be stable, so that we can think of our theory as a standard relativistic QFT constructed about a (perturbatively) stable Poincar\'e invariant vacuum state, then at item 2 above we need $c_2 > 0$. 
This is compatible with everything else, however.

%%%%%%%%%%%%%%%%%%%%%%%%%%%%%%%%%%%%%%%%%%%%%%%%%%%%%%%%%
%%%%%%%%%%%%%%%%%%%%%%%%%%%%%%%%%%%%%%%%%%%%%%%%%%%%%%%%%
\section{The quantum theory}\label{amplitude}

So far we focused on classical aspects of our solution. However we also have to impose that it make sense quantum-mechanically---that is, that it be within the regime of validity of our effective theory. At first sight, this is not difficult to achieve. The constraints derived in the previous section are linear in $c_n H_0^{2(n-2)}$, and thus can be generically satisfied by choosing
%For the deSitter configuration (\ref{dS_solution}) to be a solution, the coefficients of the non-linear terms in the Lagrangian must be weighed by suitable powers of $H_0$. Namely, working with a dimensionless $\pi$ like we have been so far, we have
\be \label{c_n}
c_2 \equiv f^2 \; , \qquad c_{n > 2} =a_n\frac{f^2}{H_0 ^{2(n-2)}} \; .
\ee
with arbitrary $f^2$ and with $a_n=O(1)$.
Now, $f^2$ appears as an overall multiplicative constant in the Lagrangian, and therefore its value cannot affect the classical solutions. On the other hand, quantum-mechanically it formally plays the role of $1/ \hbar$. The bigger $f^2$, the smaller the quantum mechanical effects. Indeed this is made manifest by going to canonical normalization for $\pi$: $\pi_c \equiv f \, \pi$. Then all self-interactions for $\pi_c$ are suppressed by powers of $H_0$ {\em and} of $f$.
By keeping $H_0$ fixed and taking $f \gg H_0$, we can make quantum effects arbitrarily weak while keeping a given classical solution unchanged.

%This means that for $f \gg H_0$, the strong-coupling energy-scale of the theory is parameterically higher than $H_0$, which is the typical gradient scale of the solution.

Still, the limit $f \gg H_0$ entails not quite an inconsistency---but a peculiarity of our model.  The problem lies in the structure of the coefficients of $2 \to 2$ scattering amplitudes at low energies, as we now explain. Our analysis is a refinement, in a quantitative way, of the argument presented in ref.~\cite{AADNR}. In that paper it has been pointed out, that for an effective theory to be the low-energy limit of a relativistic microscopic theory obeying the standard properties of the $S$-matrix (analyticity, unitarity, Froissart bound), the $2 \to 2$ forward scattering amplitude for identical particles cannot go to zero faster than $s^2$ at low energies---and the coefficient of the $s^2$ term has to be positive. Now, the galileon grossly violates this, before the conformal completion we adopted. This is because each galilean interaction involves too many derivatives per field (see eqs.~(\ref{L3}, \ref{L45})). And in fact at tree-level the $2 \to 2$ amplitude vanishes exactly in the forward limit ($t \to 0$)---by power counting it should behave like $s^3$, but crossing symmetry in the forward limit implies parity under $s \to -s$. 

The conformal completion can in principle fix this, because it supplements the galileon Lagrangian with interaction terms with fewer derivatives per field \cite{NRT}. For instance with ${\cal L}_3$ there is an associated $(\di \pi)^4$ term, which has exactly the right structure to yield an $s^2$ contribution to the scattering amplitude.
On the other hand, the conformal version of ${\cal L}_4$ has too many derivatives, and that of ${\cal L}_5$ has too many $\pi$'s (as well as too many derivatives), so we can neglect them. So, the part of the Lagrangian that can contribute an $s^2$ piece  to the forward scattering amplitude is (see eqs.~(\ref{L2c}, \ref{L3c}))
\bea
{\cal L}  & \supset & c_2 {\cal L}_2 + c_3 {\cal L}_3  \nonumber \\
& = &  - \sfrac{1}{2}\,c_2 \, e^{2\pi} (\di \pi)^2 - \sfrac{1}{2}\, c_3 \, \Box \pi (\di \pi)^2 - \sfrac{1}{4} \, c_3 \,  (\di \pi)^4  \nonumber \\ 
& \to & -\sfrac12 c_2 \, (\di \tilde \pi)^2 + \sfrac14 c_3 (\di \tilde \pi)^4 + \dots \; , \label{redefine}
\eea
where we got the last line after suitable non-linear field redefinitions, and the dots stand for higher orders in $\tilde \pi$ and derivatives thereof. 

We thus see that with our conformally invariant Lagrangian, we can have a positive $s^2$ term in the low-energy forward $2\to2$ scattering amplitude,
\be \label{amp}
{\cal M}(s, t\to 0) \sim \frac{c_3}{c_2^2} s^2 + \dots \; ,
\ee
for positive $c_3$. Notice that $c_3>0$ is consistent with our `recipe' of sect.~\ref{recipe}.
The strong coupling scale associated with this term in the amplitude is 
\be
\Lambda  = \big( c_2^2 / c_3\big) ^{1/4} \sim \big( f H_0) ^{1/2} \; .
\ee
However, the strong coupling scales associated with higher-order terms can be lower. For instance the cubic interaction $\Box \pi (\di \pi)^2$ used twice in two trilinear vertices connected by a propagator yields an $s^3 +  t^3 + u^3$ term (which vanishes in the forward limit but is non-zero otherwise) suppressed by the scale
\be \label{Lambdaprime}
\Lambda' = \big( c_2^3 / c_3^2 \big) ^{1/6} \sim \big( f H_0^2) ^{1/3} \;.
\ee
This is much lower than $\Lambda$ for $f \gg H_0$. More systematically, expanding the tree-level Lagrangian in powers of the canonicallly normalized field $\pi_c$, we get interactions of the form
\be
{\cal L}_{n,m,q} \sim  \di \pi_c \di \pi_c \big( \pi_c / f \big)^n \big( \di  \pi_c / H_0 f \big)^{2m} 
\big( \di\di  \pi_c / H_0^2 f \big)^q \; ,
\ee
with $1 \le m+q \le 3$ and arbitrary $n$, except for $m=0, q=1$, in which case $n=0$. The associated strong-coupling scales are
\be
\Lambda_{n,m,q} = H_0 \big( f / H_0 \big) ^\frac{n+2m+q}{n+4m+3q} \; .
\ee
For $f \gg H_0$ the lowest of such scales corresponds to $n=m=0$,
\be
\Lambda_{\rm strong} \equiv \Lambda_{0,0,q} = H_0 \big( f / H_0 \big) ^{1/3} \; ,
\ee
which is exactly the scale $\Lambda'$ of eq.~(\ref{Lambdaprime}).

Now, the fact that the leading low-energy term in the amplitude is weighed by a scale that is much higher than that weighing higher-order terms is not inconsistent, but suggests that the UV-completion of the model, if it exists, is weakly coupled. In such a case, the different strong-coupling scales $\Lambda_{n,m,q}$ do not correspond to any physical scales in the theory---the theory never becomes strongly coupled---they are just low-energy mirages arising from taking different combinations of a weak coupling and of the true new physics mass-scale of the theory. This is evident for instance in the tree-level exchange  of a weakly coupled massive particle at low energies:
\be \label{exchange}
g^2 \frac{1}{p^2 + M^2} \to \frac{1}{\big(M/g \big)^2} - \frac{p^2}{\big(M/g^{1/2} \big)^4}  + \frac{p^4}{\big(M/g^{1/3} \big)^6}  + \dots \; ,\qquad p^2 \ll M^2 
\ee
(as we will see our case suggests a slightly subtler interpretation than this simple example).
Now, because of scale invariance (\ref{dilation}) and of eq.~(\ref{c_n}), the structure of our  Lagrangian is
\be
{\cal L}  =  f^2 \, e^{2 \pi} \, \di \pi \di \pi \,  F\big(  e^{-\pi} \di \pi / H_0,  e^{-2\pi} \di \di \pi / H_0^2 \big) \,
\ee
where the function $F$ is dimensionless, and has no small parameter. Then, it is natural to conjecture that the derivative expansion be controlled by the scale
\be
\phi(x) \equiv e^{\pi(x)} H_0 \; .
\ee
That is, if we rewrite our classical Lagrangian as
\be \label{phi}
{\cal L} = \frac{1}{g^2} \cdot \phi^4  \, \tilde F\big(  \di \phi/ \phi^2,  \di \di \phi / \phi^3) \; , \qquad g^2 \equiv H_0^2/f^2 \; ,
\ee
its form invites the interpretation of $g$ as a coupling constant---weak, in the limit $f^2 \gg H_0^2$---and of $\phi$ as a dilaton controlling the mass-scale where the UV-completion enters. If this interpretation is correct, then the weakly coupled limit $g \ll 1$ indeed has formally very high strong coupling scale, but this says nothing about where the effective theory is actually breaking down. New physics has to enter at the scale $\phi$. For the trivial solution $\pi =0$ we have $\phi = H_0$, and the effective theory is valid at energies below this. For our deSitter solution (\ref{dS_solution}) we have
\be
\phi_{\rm dS} = \frac{1}{|t|} \; ,
\ee
and the solution itself is outside the effective theory, in the sense that all terms in $\tilde F$ above are of the same order, thus signaling a breakdown of the derivative expansion.

%We want to stress that from the low-energy standpoint we are not forced to this conclusion. As stressed at the start of this section, for $f$ much bigger than $H_0$ quantum effects are small, and the effective theory is unitary up to energies of order of the lowest of the strong-coupling scales $\Lambda, \Lambda', \dots$
%Rather, the problem arises when trying to accommodate the requirement that our theory be the low-energy limit of a `standard' relativistic microscopic theory. This demands the presence of an $s^2$ term in the low-energy amplitude, which in our case appears with a very small coefficient---smaller than those of higher order terms. This {\em usually} corresponds to the UV-completion being weakly-coupled, thus implying that the effective theory in fact breaks down much before expected.

We want to stress that, in fact, we are not forced to this conclusion. If $\tilde F$ were an infinite series---the full derivative expansion---then indeed identifying the UV-completion's mass scale would be unambiguous. For instance  one could compute the $2 \to 2$ tree-level scattering amplitude at all orders in $s/\phi^2$ about a $\phi={\rm const}$ solution. Then the radius of convergence of that series---the distance of the first non-analyticities from the origin---would define the scale of new physics.  Perhaps more physically, the strong-coupling scale associated with a very high-order term would be very close to $\phi$, the extra factor of $g$ being diluted by a tiny fractional power (see eq.~(\ref{exchange})). 
In our case instead $\tilde F$ is a finite polynomial, and the identification of the UV-scale is not as straightforward.
For $g \ll 1$ quantum effects are small, and the effective theory is unitary up to energies of order of the strong-coupling scale $\Lambda_{\rm strong} = H_0/g^{1/3}$.

We can however put an upper bound on the UV-completion's scale by means of dispersion relations, and such a bound turns out to be parameterically smaller than the strong coupling scale---in agreement with the idea that the UV-completion be weakly coupled. If we assume analyticity, unitarity, and that the forward $2\to2$ scattering amplitude ${\cal A}(s) \equiv {\cal M} (s, t \to 0)$ is bounded by $s^2$ at large $s$, then we have \cite{AADNR}
\be\label{amp2}
{\cal A}''(0) = \frac{2}{\pi} \int_0^\infty \frac{\sigma(s)}{s^2} ds 
\ee
where $\sigma(s)$ is the total cross-section for $\pi\pi$ scattering.
With our effective theory we can compute the left-hand side, and the low-energy contribution to the right-hand one. Since the integrand is positive, this equation then yields an upper bound $\Lambda_{\rm UV}$ on the scale where we should stop trusting our low-energy computations---the cutoff of the effective theory. 
About the trivial solution $\pi = 0$, $\phi=H_0$, the l.h.s.~can be got from eq.~(\ref{amp}): 
\be
{\cal A}''(0) \sim \frac{c_3}{c_2^2} = \frac{g^2}{H_0^4} \; .
\ee
On the other hand the low-energy cross-section is UV-dominated (as befits a derivatively coupled theory), and in particular at energies above $\Lambda_{0,0,1}^3/\Lambda_{0,1,0}^2 = H_0$ it is dominated by the $s^3$ piece in the amplitude, thus scaling as
\be
\sigma(s) \sim g^4 \frac{s^5}{H_0^{12}} \; . 
\ee
We thus see that in order not to violate eq.~(\ref{amp2}), the low-energy contribution has to be cut off below
\be \label{LambdaUV}
\Lambda_{\rm UV} = \frac{H_0}{g^{1/4}} \; .
\ee
This defines the highest possible energy scale our effective theory can be extrapolated to (under the assumptions given above). The UV-completion has to enter below this scale. Notice that $\Lambda_{\rm UV}$ is parameterically higher than $\phi = H_0$, which was the cutoff scale `naturally' suggested by the structure of eq.~(\ref{phi}). This makes our deSitter-like solution consistent with the standard $S$-matrix properties of well behaved relativistic theories. Notice also that $\Lambda_{\rm UV}$ is lower than the strong-coupling scale,
\be
\Lambda_{\rm strong} = \frac{H_0}{g^{1/3}} \; ,
\ee
as it should be for the consistency of our estimate.

In general there may be infinitely many more dispersion relations that can constrain further the structure of the low-energy effective Lagrangian, and set an upper-bound on its cutoff lower than eq.~(\ref{LambdaUV}). For instance one could consider $n \to n$ scattering amplitudes with $n$ larger than two.
Such constraints are going to be quite cumbersome to work out, and are beyond the scope of the present paper.
On the other hand, in the case of weakly coupled UV-completion further information can be extracted from the forward $2\to2$ amplitude ${\cal A}(s)$ \cite{AADNR}. We will not take this route either---in fact it is straightforward to show that in our case nothing new is learned on the cutoff of the theory by taking it. Instead, we will consider signal propagation in the presence of weak background fields. This will turn out to be surprisingly powerful.

\section{Retaining causality}

Ref.~\cite{AADNR} identified a class of derivatively coupled theories for which $S$-matrix positivity along the lines of last section implies the sub-luminality of excitations in the presence of non-trivial background fields. In our case, as we will see in a moment, quite the opposite is true: amplitude positivity forces $c_3$ to be positive, which in turn makes certain excitations about background fields {\em super}-luminal, unless the cutoff of the theory is low enough. We will use this to constrain the cutoff of our effective theory. Notice that the analyses of refs.~\cite{AADNR, NRT} already exhibited the presence of super-luminal excitations for DGP-like and galileon theories about certain background solutions. We find it useful however to provide a self-contained similar analysis here, mainly for two reasons. The first is that we will exhibit superluminality for essentially {\em any} static weak-field solution to the field equations---which eliminates the possibility that perhaps superluminality about a particular solution just signals that the field cannot be coupled to sources in such a way as to produce that solution. The second reason  is that we will show that superluminality is gone if  the UV cutoff of the theory is of order $\langle \phi \rangle$.

To study the propagation of fluctuations about a background solution, it is more convenient to use $\phi=H_0 e^\pi$  
as field variable. The Lagrangian becomes (see  eq.~(\ref{redefine}))
\be \label{L_phi}
{\cal L} = \frac{1}{g^2} \bigg[ -\frac12 (\di \phi)^2 - \frac{\tilde c_3}{2} \, \frac{\Box \phi (\di \phi)^2}{\phi^3} 
+ \frac{\tilde c_3}{4} \frac{(\di \phi)^4}{\phi^4} + \dots \bigg] \; ,
\ee
where $g =  H_0/f $, $\tilde c_3 \equiv c_3 H_0^2/f^2$, and the dots stand for higher-order terms,  coming from ${\cal L}_4$ and ${\cal L}_5$. As clear from last section's discussion, the dimensionless coefficient $\tilde c_3$ has to be positive, and of order unity. Now, consider a background solution $\phi_0(x)$, and small fluctuations $\varphi$ about it. At quadratic order the fluctuation Lagrangian is
\bea
{\cal L}_\varphi   & = & \frac{1}{g^2} \bigg[  -\frac12 (\di \varphi)^2 \Big(1 + \tilde c_3 \frac{\Box \phi_0}{\phi_0^3} -  \tilde c_3 \frac{(\di \phi_0)^2}{\phi_0^4} \Big) + \tilde c_3 \Big( \frac{\di^\mu \di^\nu \phi_0}{\phi_0^3} - 2 \frac{\di^\mu \phi_0 \di^\nu \phi_0}{\phi_0^4} \Big) \di_\mu \varphi \di_\nu \varphi + \dots \bigg ]  \nonumber \\
& \equiv & -  \frac{1}{2 g^2} \,   G^{\mu\nu} \, \di_\mu \varphi \di_\nu \varphi + \dots \; , \label{kinetic}
\eea
where the dots stand for non-derivative terms (i.e.~mass terms) for $\varphi$, as well as for contributions from the terms already neglected in eq.~(\ref{L_phi}), and we defined the effective inverse metric $G^{\mu\nu}$  as
\be \label{G}
G^{\mu\nu} = \eta^{\mu\nu} \Big(1 + \tilde c_3 \frac{\Box \phi_0}{\phi_0^3}  - \tilde c_3 \frac{(\di \phi_0)^2}{\phi_0^4} \Big) -2  \tilde c_3 \Big( \frac{\di^\mu \di^\nu \phi_0}{\phi_0^3} - 2 \frac{\di^\mu \phi_0 \di^\nu \phi_0}{\phi_0^4} \Big) 
\ee
The reason why we focus on the kinetic action for $\varphi$ and neglect non-derivative terms, is that the causal structure is defined by the highest-derivative terms: according to Leray's theorem \cite{wald}, for a Lagrangian of the form (\ref{kinetic}), $\varphi$ excitations travel along the light-cone of the effective metric $G_{\mu\nu}$.

We are thus led to analyze the light-cone of the effective metric (\ref{G}) in the presence of a non-trivial $\phi_0(x)$.
To be as general as possible, we consider a generic weak vacuum solution $\phi_0(x)$. By `weak' we mean that the non-linearities of our action are negligible. Also, for simplicity we restrict to a static configuration, so that the solution obeys
\be \label{Laplace}
\nabla^2 \phi_0 \simeq 0 \; .
\ee
The physical situation we have in mind is that we have stationary, localized sources for $\phi$, and that away from the sources the field is in its linear regime. We generically expect there to be a stationary point for $\phi_0$ 
\footnote{This is guaranteed if the source arrangement is minimally inventive. Also such an assumption is not necessary to proceed, but makes the discussion neater. Indeed a space gradient term, corresponding to a term linear in $x^j$ in eq.~(\ref{solution}), gives just a negative definite contribution  to  $G_{ij}$ in eq.~(\ref{Glower}), making the light cone even wider.}.
We zoom in on that point, and choose it as the origin of our coordinate system. The solution can be Taylor expanded about the origin
\be \label{solution}
\phi_0 (\vec x) \simeq \Phi_0 \Big( 1 + \frac12 A_{ij} \frac{x^i x^j}{L^2} \Big) \; ,
\ee
where $\Phi_0$ is the field value at the origin, $L$ is the typical scale over which the solution varies, and $A_{ij}$ is a dimensionless matrix with entries of order unity and which, because of Laplace's equation (\ref{Laplace}), is traceless. As we will see, this last property is crucial for our argument.

Now, to impose, as we assumed,  that  the solution be in a linear regime, we have to demand
\be
L \gg 1 / \Phi_0 \; .
\ee
Indeed the size of non-linearities in eq.~(\ref{L_phi}) for the terms explicitly kept is of order $\di^2/\phi^2$, which for a configuration like (\ref{solution}) is precisely $1/(L \Phi_0)^2$. The neglected terms in eq.~(\ref{L_phi}) are suppressed by even higher powers of $\di^2/\phi^2 \sim 1/(L \Phi_0)^2$---that is why we are allowed to neglect them in the first place.
Not surprisingly, the size of non-linearities $1/(L \Phi_0)^2$ is also controlling the corrections to $\eta^{\mu\nu}$ in the inverse metric (\ref{G}). As long as this is small, the latter can be straightforwardly inverted to yield the effective metric
\be \label{Glower}
G_{\mu\nu} \simeq \eta_{\mu\nu}  + 2  \tilde c_3 \bigg( \frac{\di_\mu \di_\nu \phi_0}{\phi_0^3} - 2 \frac{\di_\mu \phi_0 \di_\nu \phi_0}{\phi_0^4} \bigg) \; ,
\ee
where we neglected the correction proportional to $\eta_{\mu\nu}$, because it does not affect the light-cone aperture. That is, at this order we can factor it out of the metric, as an overall conformal factor, which is irrelevant for our question.

To check whether the light-cone of $G_{\mu\nu}$ includes directions that are super-luminal with respect to the true Minkwoski light-cone defined by $\eta_{\mu\nu}$, we contract $G_{\mu\nu}$ with a generic null direction $n^\mu = (1, \hat n)$. If $n^\mu$ turns out to be time-like for  $G_{\mu\nu}$ at least for some direction $\hat n$, we know that $\varphi$ signals can travel superluminally along $\hat n$. For the moment we consider small distances from the origin, $x \ll L$, so that the denominators in (\ref{Glower}) can be taken as constant for our solution.
We have
\be
G_{\mu\nu} \, n^\mu n^\nu \simeq  \frac{2 \tilde c_3}{(L \Phi_0)^2} \bigg[ A_{ij}  -2  \frac{(A_{ik} x^k)(A_{jl} x^l)}{L^2} \bigg] \hat n^i \hat n^j \; .
\ee
The second term in brackets is always a negative definite matrix, being minus the tensor product of $A \cdot \vec x$ with itself. It thus gives a superluminal contribution in all directions---that is, it tends to make $n^\mu$ time-like w.r.t.~$G_{\mu\nu}$ regardless of $\hat n$.
However at small distances from the origin, $x \ll L$, the second term is negligible with respect to the first. Now, $A_{ij}$, being traceless, has at least one negative eigenvalue. The corresponding eigenvector, defines a direction $\hat n$ for which the first term also gives a superluminal shift in the propagation speed, of order
\be \label{superluminal}
\Delta c \sim \frac1{(L \Phi_0)^2} 
\ee
(recall that both $\tilde c_3$ and $A_{ij}$ are of order one.)
We are thus led to the conclusion that a generic linear solution $\phi_0(x)$ allows for superluminal excitations, because the effective metric perturbations propagate on has a light-cone that is broader than that defined by $\eta_{\mu\nu}$. We stress that this result  does not depend on the detailed form of the coupling of $\phi$ to each individual source. For instance on whether it couples through a `monopole'   $\phi J$, or via dipole $\di_\mu \phi  J^\mu$ or even via higher multipoles. This generalizes the proof  of  superluminality given in ref.~\cite{NRT} by considering spherical solutions induced by monopole coupling.

The one we just obtained is a purely classical result: at the classical level, to ascertain the existence of superluminal signals, it is enough to analyze the effective metric (Leray's theorem). Upon canonical quantization the causal structure does not change \cite{DNTV}, but since we have an effective theory with a cutoff, to be consistent we have to demand that the superluminal effect we isolated be measurable within the effective theory. The superluminal shift (\ref{superluminal}) in the propagation speed entails superluminal propagation of actual signals if, for a mode of given momentum $k$ propagating from the origin all the way to $L$, the gain over a strictly luminal signal is at least of order of one wavelength. That is, we need
\be
\Delta c\,  kL \sim \frac{k}{\Phi_0} \frac{1}{L \Phi_0} \gtrsim 1 \; .
\ee
We are restricting to a region of size $L$ around the origin, because that is where our approximations hold. We will comment below on what happens outside this region. Recall that for the solution to be linear, we want $L \Phi_0$ larger than one. We thus see that the superluminal effect is unobservable if we restrict to fluctuations with momenta smaller than $\Phi_0$. Therefore, if we postulate that the cutoff of our effective theory is the local value of $\phi$, like the structure of our Lagrangian (\ref{phi}) naturally suggested, we cannot consistently conclude that the causal structure of the theory is different than the usual relativistic one.

Nothing new is learned by relaxing the $x \ll L$ restriction. For a generic linear solution, with a generic arrangement of sources, the structure of the solution will deviate significantly from eq.~(\ref{solution}) beyond $x \sim L$, and full knowledge of the solution will be needed to conclude anything about signal propagation. However since eq.~(\ref{solution}) is {\em a} good vacuum solution for $x \gg L$ as well---non-linear effects are down by powers of $L^4/(\Phi_0^2 x^6)$---one could study how signals propagate on it for distances much longer than $L$. The problem is that the superluminal effect is proportional to the leading non-linearity, and so the former also rapidly  goes to zero at very large $x$. That is, for a solution like eq.~(\ref{solution}) the superluminal effect is concentrated in a region of size $L$ around the origin.

In summary, if we assume that our effective theory arises as the low-energy limit of a microscopic theory with the standard relativistic causal structure, we are {\em forced} to bound its regime of validity by the local value of $\phi$. Under those assumptions our stable NEC-violating solution of sect.~\ref{NECsolution} is therefore untenable.

%%%%%%%%%%%%%%%%%%%%%%%%%%%%%%%%%%%%%%%%%%%%%%%%%%%%%%%%%
%%%%%%%%%%%%%%%%%%%%%%%%%%%%%%%%%%%%%%%%%%%%%%%%%%%%%%%%%
\section{Back to the amplitude}

In hindsight, it is natural to ask whether the $S$-matrix encodes  any signs of the superluminal effect we unveiled. In simpler examples, like those considered in ref.~\cite{AADNR}, sub-luminality of excitations about background fields is tied to positivity of forward $2 \to 2$ amplitudes at order $s^2$. In our case instead, as we discussed, such positivity is obeyed so long as $c_3$ is positive, yet we have super-luminality regardless of $c_3$'s sign. In fact, our super-luminal effect relies on the cubic interaction $\Box \pi (\di \pi)^2$ whose contributes to $2 \to 2$ amplitudes is of order $s^3$ and vanishes in the forward limit, being proportional to $s^3+t^3+u^3$. It thus looks like that in order to isolate the term responsible for super-luminality, we need to consider non-forward amplitudes.

Some form of positivity still holds for non-forward $2\to2$ amplitudes ${\cal M}(s,t)$. Indeed following Martin\footnote{For a recent application see also \cite{Vecchi:2007na}.} \cite{Martin}, unitarity together with standard properties of Legendre polynomials imply that for real $s$
\be \label{dMdt}
\frac{\di^n}{\di t^n} {\rm Im} \, {\cal M}(s + i \epsilon, 0) \ge 0 \qquad \forall n \;, 
\ee
which is a generalization of the optical theorem, $ {\rm Im} \, {\cal M}(s+ i \epsilon, 0) \propto \sigma_{\rm tot} > 0 $. 
{\em If} there is a mass-gap, ${\cal M}(s + i \epsilon, t)$ is analytic in the $t$-plane, for $|t|$ below threshold\footnote{More precisely, as proven in ref. \cite{Martin},   analyticity in $(s,t)$ holds on the product space $(|t|<M) \times ({\rm cut}\, s\, {\rm plane})$.}. This means that we can Taylor-expand ${\cal M}(s + i \epsilon, t)$ in $t $ about the origin, and use the inequalities (\ref{dMdt}) to conclude
\be
{\rm Im} \, {\cal M}(s + i \epsilon, t) > 0 \qquad {\rm for}  \; t > 0 \; ,
\ee
so long as $t$ (now real and positive) is below threshold. We can then run a standard dispersion argument (see e.g.~\cite{AADNR})---this time using analiticity in the $s$-plane---and show that
\be \label{positivity}
\frac{\di^2}{\di s^2} {\cal M}(0, t)  = \frac{2}{\pi} \int \! d s \,{\rm Im} \, {\cal M}(s + i \epsilon, t ) \bigg[ \frac{1}{s^3} + \frac{1}{(s+t)^3} \bigg] > 0  \qquad {\rm for}  \; t > 0 \; ,
\ee
where the integral is taken along the cut on the positive real axis, from the threshold to infinity.

Now, in our case there is no mass-gap---the particles we are scattering are themselves massless---and we have non-analyticities at arbitrarily low $s$ and $t$. However, we already argued that the UV-completion of our model must be weakly coupled, with coupling $\lambda \ll 1$. We can then consider the scattering amplitude at tree-level in $\lambda$. Such an amplitude has non-analiticities associated with the exchange of the UV-completion's massive states, but it does not have low-energy cuts associated with multi-massless particle states---these enter at higher order in $\lambda$ \cite{AADNR}. It thus makes sense to apply the above dispersion relation to the tree-level amplitude, even in the absence of a mass gap. The mass-scale playing the role of the gap is then the UV cutoff of the effective theory, where the UV-completion enters and the first non-analiticities show up.

Now, in our case we assumed that the cutoff of the theory is parameterically higher than $H_0$, perhaps as high as eq.~(\ref{LambdaUV}). At intermediate energies, $H_0 \ll E \ll \Lambda_{\rm UV}$, the amplitude is dominated by the ${\cal O}(s^3)$ terms, and we can neglect the ${\cal O}(s^2)$ ones. The former gets a contribution from the Lagrangian term we are trying to isolate, $\Box \pi (\di \pi)^2$, but also directly from the quartic galilean interaction contained in ${\cal L}_4$. Indeed we can perform a non-linear field redefinition that eliminates the cubic interaction and yields a quartic term of the form
\be
\di \pi \di \pi \, \di \di \pi \, \di \di \pi \; ,
\ee
which is precisely the structure of ${\cal L}_4$. Therefore, unless the ${\cal L}_4$ contribution accidentally vanishes, which it does not, we already see that our attempt to connect super-luminality with violation of amplitude positivity, fails: amplitude positivity can be fixed by a suitable choice of $c_4$, which as we argued does not affect signal propagation in the presence of weak background fields. However, if we want our Lagrangian to describe NEC-violating stable solutions, $c_4$ is not completely arbitrary. It may be that once we impose the conditions of sect.~\ref{recipe}, positivity turns out to be violated. 

Computing the ${\cal O}(s^3)$ tree-level amplitude is straightforward. The result is
\be
{\cal M}(s,t)  = \frac34 \frac{1}{c_2^3} \big(c_3^2 - 2 \, c_2 c_4 \big) s t (s+t) \; .
\ee
Positivity in the form of eq.~(\ref{positivity}) is obeyed iff
\be
c_3^2 > 2 \, c_2  c_4 \; .
\ee 
Such an inequality is in fact {\em implied } by our conditions of sect.~\ref{recipe}, as manifest from the following simple algebraic manipulation:
\be
 2c_2 c_4 < 4(c_3 c_2 - c_2^2 ) \le c_3^2 \; ,
\ee
where in the first step we used eq.~(\ref{c4}) and that $c_2 > 0 $ for the stability of the $\pi=0 $ vacuum, and in the last step
we extremized over $c_2$. 
In conclusion, by assuming that our effective theory can be extrapolated to energies parameterically higher than $H_0$ we find no contradiction with amplitude positivity, even in the non-forward case.
Conversely, absence of superluminal excitations demands the theory to be cut off at $H_0$.

%%%%%%%%%%%%%%%%%%%%%%%%%%%%%%%%%%%%%%%%%%%%%%%%%%%%%%%%%
%%%%%%%%%%%%%%%%%%%%%%%%%%%%%%%%%%%%%%%%%%%%%%%%%%%%%%%%%
\section{Discussion}

Our effective theory allows for an isotropic NEC-violating solution that is stable against small fluctuations. Moreover, signal propagation on this solution is exactly luminal. This is a nice counterexample to the theorem of ref.~\cite{DGNR}, which here is violated because one of the hypotheses, subtly, is. The relevant part of the Lagrangian formally involves higher-derivative terms, yet the corresponding field equations are second-order.

Still, if the goal was to find a completely sensible relativistic QFT that can violate the NEC in the semi-classical regime, then we failed. From the standard EFT standpoint, our theory {\em is} perfectly sensible, and the relevant NEC solution is within its regime of validity. Moreover, our theory obeys standard positivity of $2\to2$ amplitudes, even in the non-forward case---which is a stronger requirement than purely low-energy consistency \cite{AADNR}. 
Nevertheless, if we start from the vacuum and excite a generic weak field configuration, small excitations about this background will now propagate super-luminally.
The presence of super-luminal excitations signals that the causal structure of the theory is not the standard Lorentz-invariant one. As a consequence our effective theory cannot arise as the low energy limit of a Lorentz-invariant UV-complete local QFT, for which micro-causality has to hold as an exact (i.e., non-perturbative) operator statement \cite{AADNR, DNTV}.

Now, we can avoid super-luminality if the cutoff of our theory is in fact much lower than expected. However for such a low cutoff, our NEC-violating solution is outside the effective theory and thus cannot be trusted.
That is: either {\em (i)} we believe that the cutoff is high enough to make room for the NEC-violating solution, but then we have to accept the existence of super-luminal excitations about other solutions, also inside the effective theory; or {\em (ii)}
we believe that the cutoff is low enough to forbid superluminal propagation of signals, but then our NEC-violating solution is gone as well. Essentially, absence of superluminal excitations demands that all non-linearities be small: if the cutoff of the theory is the local value of $\phi$, as required by sub-luminality, then all non-linear terms in eq.~(\ref{phi}) have to be small for the solution to be within the effective theory, despite what suggested by purely low-energy EFT considerations and even by dispersion relations. This makes the tension we found particularly robust. It is not an accident of our particular solution, but rather it is a property of any solutions that crucially rely on non-linearities. Since the NEC is obeyed by the free part of the Lagrangian, {\em all} NEC-violating classical solutions are outside the regime of the effective theory. 

We thus see that there still is a strong tie linking the NEC and subluminality. It is less direct than in the absence of higher-derivative terms, where it is the NEC-violating solutions themselves that support superluminal excitations \cite{DGNR}, but it appears to be equally tight. It would be interesting however to understand whether this relation is an accident of our Lagrangian, or whether instead it is a more general, and thus deeper, fact.
For instance, we adopted a conformal modification of the galileon theory in order to respect positivity at order $s^2$ in the scattering amplitude. However this requirement by itself does not single out the conformal group $SO(4,2)$ as the symmetry group of our theory. We chose it because then with such a high degree of symmetry, finding very symmetric solutions and studying their properties is immediate. However there are probably other equally convenient choices---for instance the 5D Poincar\'e group $ISO(4,1)$ \cite{NRT}---and in principle nothing stops us from lowering the number of symmetry generators, as long as galilean invariance is an approximate symmetry in the UV. Note that for superluminality to happen, one just needs a non-vanishing galilean cubic term. In our theory $c_3$ has to be strictly positive for the NEC-violating solution to be well behaved (sect.~\ref{recipe}) and for the forward amplitude to be positive at order $s^2$ (sect.~\ref{amplitude}). It is entirely possible that disentangling the coefficient of the cubic galilean term from others' could eliminate superluminality altogether while retaining the good features of our theory as well as of the NEC-violating solution.

From a more general field theoretical perspective, we find it interesting that in our theory relativistic dispersion relations involving 
the $2 \to 2$ scattering amplitude are not enough to discover superluminality in the presence of sources. 
In the simple cases of ref.~\cite{AADNR}, the link was so direct that forward dispersion relations were enough to imply sub-luminality about any background solution. In our case instead, even {\em non-forward} ones are obeyed by a theory with superluminal excitations. We expect that at some level of complexity (for instance one could consider $n \to n $ scattering amplitudes, with $n>2$) dispersion relations should encompass the requirement of subluminality about any background fields. Indeed the sources turning on these backgrounds may be viewed as fictitious ones---used to construct the generating functional $W[J]$ from which the $n$-point functions $\langle \phi(x_1) \dots \phi(x_n) \rangle$
are then straightforwardly derived. It is clear then, that the propagation properties of $\phi$ fluctuations about some background field {\em are} encoded in these $n$-point functions, and thus in the $S$-matrix elements. It would be interesting to understand for a generic theory the $S$-matrix equivalent of the sub-luminality requirement.

\section*{Acknowledgements} 
We would like to thank Justin Khoury for spotting a mistake in eq.~\eqref{culprit} in the first version of our paper, and for discussions. We would also like to thank Paolo Creminelli for discussions and for joining us in checking our algebra for this second version of the paper.
A.N.~and E.T.~would like to thank the Institute of Theoretical Physics of the Ecole Polytechnique F\'ed\'erale de Lausanne for hospitality during this project. The work of A.N. is supported by NASA ATP (09-ATP09-0049). The work of R.R.~is partially supported by the Swiss National 
Science Foundation under contract No.~200021-116372 and 200021-125237.

%%%%%%%%%%%%%%%%%%%%%%%%%%%%%%%%%%%%%%%%%%%%%%%%%%%%%%%%%
%%%%%%%%%%%%%%%%%%%%%%%%%%%%%%%%%%%%%%%%%%%%%%%%%%%%%%%%%

\end{document}